\documentclass[twocolumn, floatfix]{aastex701}
\usepackage[dvipsnames]{xcolor}
\usepackage{amsmath, comment, natbib}
\usepackage{array} 
\begin{document}

\title{Photon‑Count Statistics of Crab X‑ray Pulses: Skellam Behavior and Excess Variance in the Main Pulse}

\author{Max Worchel}
\email[show]{mworchel@haverford.edu}
\NoHyper\correspondingauthor{Max Worchel}\endNoHyper
\affiliation{Haverford College Department of Physics and Astronomy, 370 Lancaster Ave, Haverford, PA 19041, USA} 

\author{Margaret M.Ferris}
\email{mmferris@haverford.edu}
\affiliation{Haverford College Department of Physics and Astronomy, 370 Lancaster Ave, Haverford, PA 19041, USA} 

\author{Sasha Levina}
\altaffiliation{Current Address: Department of Astronomy \& Astrophysics, University of California San Diego, La Jolla, CA 92093, USA}
\email{slevina@ucsd.edu}
\affiliation{Haverford College Department of Physics and Astronomy, 370 Lancaster Ave, Haverford, PA 19041, USA} 

\author{Iris Horn}
\altaffiliation{Current Address: Cobbs Creek Healthcare, Newtown Square, PA 19073}
\email{iris.horn8@gmail.com}
\affiliation{Haverford College Department of Physics and Astronomy, 370 Lancaster Ave, Haverford, PA 19041, USA} 

\author{Mac Tygh}
\altaffiliation{Current Address: Johns Hopkins University Department of Italian, MD 21218, USA}
\email{mtygh1@jhu.edu}
\affiliation{Haverford College Department of Physics and Astronomy, 370 Lancaster Ave, Haverford, PA 19041, USA} 

\author[0000-0003-4137-7536]{Andrea N. Lommen}
\email{alommen@haverford.edu}
\affiliation{Haverford College Department of Physics and Astronomy, 370 Lancaster Ave, Haverford, PA 19041, USA} 

\author{Kent S. Wood}
\email{kentswood@gmail.com}
\affiliation{Praxis, resident at the Naval Research Laboratory, Washington, DC 20375, USA}

\author[0000-0002-5297-5278]{Paul S. Ray}
\email{paul.ray@nrl.navy.mil}
\affiliation{U.S. Naval Research Laboratory, Washington, DC 20375, USA}

\author{Julia S. Deneva}
\email{julia.deneva@gmail.com}
\affiliation{George Mason University,
resident at Naval Research Laboratory, Washington, DC 20375, USA}

\author{Natalia Lewandowska}
\email{natalia.lewandowska@oswego.edu}
\affiliation{State University of New York, at Oswego, Department of Physics, 7060 NY-104, Oswego, NY 13126, USA}

\author[0000-0002-0893-4073]{Matthew Kerr}
\email{matthew.kerr@nrl.navy.mil}
\affiliation{U.S. Naval Research Laboratory, Washington, DC 20375, USA}

\author[0000-0003-2742-3321]{Jeffrey S. Hazboun}
\email{jeffrey.hazboun@oregonstate.edu}
\affiliation{University of Washington Bothell, 18115 Campus Way NE, Bothell, WA 98011}

\author{David A. Howe}
\email{david.howe@nist.gov}
\affiliation{Time and Frequency Division, NIST Boulder, CO 80305, USA}
\affiliation{Department of Physics, University of Colorado Boulder, CO 80309, USA}

\author{Zaven Arzoumanian}
\email{zaven.arzoumanian@nasa.gov}
\affiliation{Astrophysics Science Division, 
NASA Goddard Space Flight Center, Greenbelt, MD 20771, USA}

\author[0000-0002-9870-2742]{Slavko Bogdanov}
\email{slavko@astro.columbia.edu}
\affiliation{Columbia Astrophysics Laboratory, Columbia University,
550 West 120th Street, New York, NY 10027, USA}

\author{Craig B. Markwardt}
\email{Craig.Markwardt@nasa.gov}
\affiliation{Astrophysics Science Division, 
NASA Goddard Space Flight Center, Greenbelt, MD 20771, USA}

\author[0000-0003-1244-3100]{Teruaki Enoto}
\email{enoto@kusastro.kyoto-u.ac.jp}
\affiliation{Department of Physics, Kyoto University, Kitashirakawa-Oiwake-cho, Sakyo-ku, Kyoto, 606-8502, Japan}

\author{Keith C. Gendreau}
\email{keith.c.gendreau@nasa.gov}
\affiliation{Astrophysics Science Division, 
NASA Goddard Space Flight Center, Greenbelt, MD 20771, USA}

\begin{abstract}

The Crab pulsar (PSR B0531+21) provides an unusually rich test bed for statistical studies of high--energy photon--counting data, owing to its extreme brightness and the contrasting behavior of its main pulse (MP) and interpulse (IP) components. Using 78.8 ks of Neutron star Interior Composition Explorer (NICER; \citealt{Gendreau2017}) data—over two million individual X--ray pulses—we construct the single--pulse photon--count distributions of the MP and IP at keV energies. We find that the IP is well described by the Skellam distribution expected for the difference of two Poisson processes, providing a rare, high--statistics empirical demonstration of Skellam behavior in an astrophysical photon--counting context. The MP also shows pulse--by--pulse variability best described by a Skellam framework when compared to Gaussian alternatives, but exhibits a significant excess variance driven by high--count events. When photon counts are summed over successive pulses, this excess averages out and the MP distribution becomes consistent with Skellam expectations, indicating that the enhanced variability does not persist across rotations. We further search for short--lag (``memory'') correlations between successive X--ray pulses and find no statistically significant lag--1 correlation. Although giant radio pulses occur in the MP phase window, their contribution is insufficient to account for the observed excess variability. Together, these results highlight a clear statistical distinction between the MP and IP and underscore the importance of using statistically appropriate models for high--energy photon--counting analyses. The distributional fits and memory limits reported here provide quantitative constraints on pulsar emission models and illustrate the broader utility of Skellam--based approaches.
\end{abstract}

\keywords{Poisson distribution (1898), Astrostatistics distributions (1884), Astrostatistics techniques (1886), Pulsars (1306), Rotation powered pulsars (1408), High energy astrophysics (739)}

\section{Introduction}

The Crab pulsar (PSR B0531+21) is an extremely bright, young pulsar located at the center of the Crab Nebula, with a spin period of 33 ms, a spin‑down rate of $\dot{P} = 4.2 \times 10^{-13}~\text{s}~\text{s}^{-1}$, and an energy‑loss power of $\dot{E} = 5 \times 10^{38}~ \text{ erg}~\text{s}^{-1}$. Since its discovery, it has been observed across the electromagnetic spectrum from radio to TeV gamma rays, yet a comprehensive description of its emission across all wavelengths has not been established \citep{Yan2022}. The Crab pulse profile exhibits multiple distinct components whose relative phases and shapes depend strongly on observing frequency. At most wavelengths, the dominant features are a main pulse (MP) and an interpulse (IP), separated by approximately 0.4 in rotational phase and connected by bridge emission at higher energies, though neither the relative alignment nor the detailed morphology of these components is constant across bands \citep{Abdo2010}.

The Crab is bright enough that pulse‑to‑pulse variability can be studied directly in radio, optical, and X‑ray regimes. Single‑pulse studies are most mature in the radio, where the high signal‑to‑noise ratio enables detailed characterization of pulse statistics and short‑duration phenomena (e.g. Hankins et al. 2016)\nocite{Hankins2016}. At higher energies, analogous studies are observationally challenging, requiring both a large effective collecting area and long cumulative exposure times. When such data are available, pulse‑to‑pulse variability provides a sensitive probe of differences between pulse components and a stringent test of statistical models applied to photon‑counting data.

In the radio band, the MP and low‑frequency IP exhibit nanoshot emission ---bursts with durations of order nanoseconds --- while the high‑frequency IP displays narrow ($<1$ GHz) emission bands at regular frequency intervals \citep{Hankins2016}. The Crab pulsar also produces giant radio pulses (GRPs), consisting of nanosecond‑ to microsecond‑duration bursts with brightness temperatures reaching up to $10^{37}$K \citep{Knight2006}. Although GRPs can occur at all rotational phases, they are most readily observed within the MP and IP phase windows \citep{Hankins2007}. These phenomena have motivated searches for correlated variability at higher energies. The peak intensity distribution of the GRPs was found to be described by a power-law function with an index of 3.3 \citep{Lundgren95}, and the peak flux was anti-correlated with GRP pulse duration \citep{Popov2007}. Their results are consistent with those of \citet{Majid2011} and \citet{Bhat2008} (power-law exponent of 2.2-2.3). Regular radio pulses have an exponential \citep{Hesse1974} or log-normal flux distribution \citep{Majid2011}. 

Using improved X‑ray timing from the Neutron star Interior Composition Explorer (NICER; \citealt{Gendreau2017}), \citet{Enoto2021} reported a statistically significant enhancement of the X‑ray flux coincident with GRPs, finding an increase of 3.8 $\pm$ 0.7\% in the MP but no significant enhancement in the IP. The magnitude of this enhancement is comparable to that observed at optical wavelengths \citep{Shearer2003, Strader2013}. These results demonstrate that pulse‑to‑pulse variability in the Crab X‑ray emission is detectable with NICER and motivate further statistical characterization of the MP and IP on a pulse‑by‑pulse basis.

Interpreting pulse‑to‑pulse variability in X‑ray data requires careful attention to the statistical properties of differenced photon counts. Although the Skellam distribution—the distribution of the difference of two Poisson variates—has been known for decades \citep{Skellam1946}, its explicit use in X‑ray astronomical data analysis has been uncommon. While we are not the first to encounter Skellam‑distributed quantities in this context, the distribution is rarely named or treated directly in the astrophysical literature. Instead, related problems involving differences of photon counts are typically addressed using Gaussian approximations or Poisson‑based constructions developed for low‑count data (e.g., Cash 1979; Gehrels 1986; Mighell 1999; Baker \& Cousins 1984)\nocite{Cash1979, Gehrels1986, Mighell1999, BakerCousins1984}. There is no widely used, “cookbook‑style” treatment of Skellam statistics tailored to astronomical photon‑counting applications. Readers interested in the mathematical background should see \citep{abramowitz1964handbook}, which in turn points back to Skellam’s original work.

While existing theoretical frameworks successfully reproduce overall energetics and average pulse profiles, they make comparatively few predictions for the statistical properties of large ensembles of individual pulses. This is an area where observational work can suggest avenues for future theoretical efforts. 

We analyze phase‑resolved X‑ray single‑pulse photon-count distributions from the Crab pulsar, modeling the main pulse (MP) and interpulse (IP) separately across millions of pulses. Our primary focus is distributional—identifying which statistical families best describe the observed amplitudes—and we also test for short‑lag correlations between successive pulses. 

The Crab pulsar presents an unusually clean and informative empirical test case for analysis via the Skellam distribution. The IP analysis establishes a well‑controlled baseline against which alternative statistical descriptions can be compared, and it provides a natural framework for testing variability in the MP.

\section{Observations}
\label{sec:observations}

We use X--ray observations of the Crab pulsar from NICER, an International Space Station (ISS) payload designed to detect X--ray photons from pulsars in the range of 0.2 to 12 keV. NICER assigns high-precision time stamps to each individual photon, which allows for precise pulsar timing which can be used to construct pulse profiles for the purpose of studying the dynamics and structure of neutron stars \citep{Gendreau2017} and for constraining the radii and the equation of state of neutron stars \citep{EOSRef}. 

The X--ray timing instrument (XTI) consists of 56 Focal Plane Modules (FPMs, 52 of which are active), which each contain a silicon drift detector \citep{Prigozhin2016}. Each FPM is paired with a grazing-incidence optic that concentrates the X--rays onto the small detector, whose geometric area is minimized in order to reduce background count rates and electron drift times. Combined, the XTI provides an effective collecting area that peaks at nearly 1,900 cm$^2$ for X--ray energies of 1,500 eV and ranges from 200 to 12,000 eV \citep{Gendreau2017}.

The background count rate depends on NICER's orbital environment and space weather conditions.  More details are available in \citet{Deneva19}. In order to increase the signal-to-noise ratio (S/N), we work with the Good Time Intervals (GTIs) containing low background. We use the standard criteria used for the processing of NICER data as detailed in \citet{Rowan2020}, including the specifications for GTIs. 

We selected observations of the Crab pulsar taken between from August 2017 to January 2021 for a total of 78.8 ks of data. The ObsIDs used in our analysis are listed in Table \ref{tab:obsids}, along with the corresponding parameter (``par") files and date ranges in MJD. Using the par files we calculate the pulse phase at each event. 
We generate par files with data from the Jodrell Monthly Crab Ephemeris\footnote{\url{http://www.jb.man.ac.uk/\~pulsar/crab.html}} \citep{parfiles}, using the jodrellcrabmonthly2par.py command from NICERsoft\footnote{See \url{https://github.com/paulray/NICERSoft/}}. 

  Each time the Crab pulsar experiences experiences a jump in rotational frequency known as a glitch, a new timing model must be used.  See \citet{timingandglitches} for a review. 

\begin{table*}

\caption{ObsIDs used, with corresponding par files}
\centering
\begin{center}
\begin{tabular}{| l | l | l |}
\hline
Par file & MJD range & ObsID(s)\\
\hline
August 2017 & 57974-58066 & 1013010108\\
November 2017 & 58067-58116 & 1013010110, 1013010113, 1013010118-1013010121\\
December 2017 & 58117-58120 & 1013010122\\
January 2018 & 58121-58189 & 1013010124\\
March 2018 & 58191-58214 & 1013010127-1013010130\\
April 2018 & 58215-58429 & 1013010133, 1013010136-1013010142\\
December 2018 & 58481-58532 & 1013010151\\
February 2019 & 58533-58724 & 1013010153, 1014020101, 2013010102, 2013010104, 2014010101\\
August 2019 & 58725-58825 & 2013010106-2013010107, 2013010109, 2205010101\\
February 2020 & 58878-58910 & 2013010112-2013010113\\
September 2020 & 59091-59123 & 3013010107-3013010108\\
November 2020 & 59152-59184 & 3013010109\\
December 2020 & 59182-59214 & 3013010110-3013010111\\
January 2021 & 59213-59245 & 3013010112\\
\hline
 \end{tabular}
 \end{center}
\label{tab:obsids}
\end{table*}

Each ObsID contains of order $10^6$ photons. The standard method of calculating the pulse phase in PINT \texttt{photonphase} uses the full pulsar model individually for each photon, but given that we have more than $10^8$ photons, the processing time became impractical.  
We therefore use polynomial coefficients to estimate the pulse phase corresponding to each photon, using the \texttt{--polycos} argument in \texttt{PINT} \texttt{photonphase}. The \texttt{polycos} argument requires input data to be barycentered, for which we use the HEASOFT barycenter correction tool \texttt{barycorr}, the DE421 JPL Ephemeris and the par files described above. Resources detailing PINT can be found in  \citet{pint1} and \citet{pint2}.

\section{Constructing Photon-Count Distributions}
\label{sec:Flux histograms}

\begin{figure}[ht]
    \centering
    \includegraphics[width=1\linewidth]{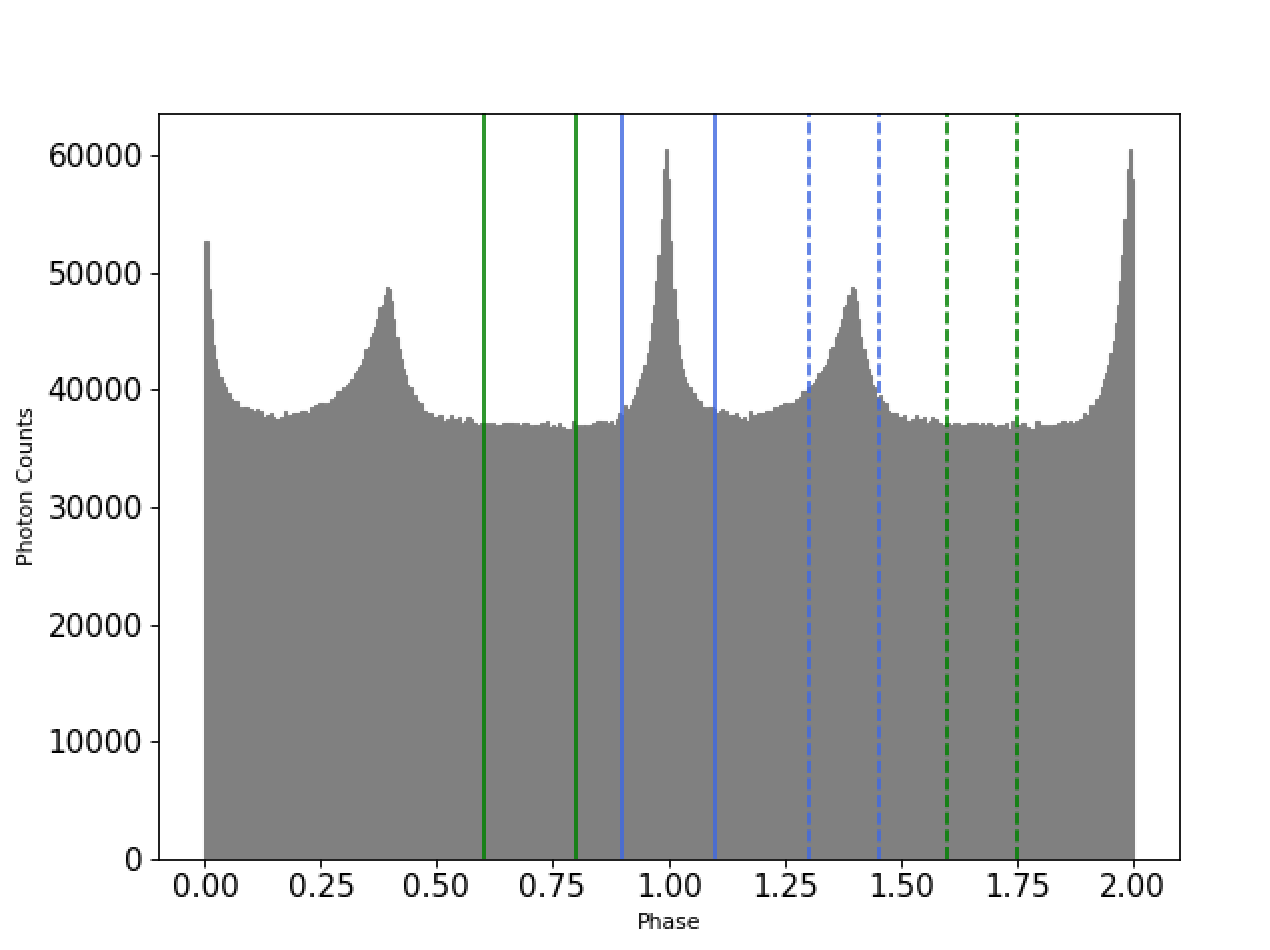}
    \caption{An example of a pulse profile from the Crab pulsar using NICER data (here with 400 bins). 
    The MP wraps around the edge of the profile, the IP is centered at 0.4 of the phase, and the bridge emission spans from the MP to the IP. The background level is approximately 37000 counts, and the pulse profile contains pulses from roughly 1,330 seconds worth of observation time. Our off-pulse windows can be seen in green, and our on-pulse windows can be seen in blue. The windows for the MP are shown in solid lines and the windows for the IP are shown in dashed lines.}
    \label{fig:pulseprofile}
\end{figure}

We define equal-width regions, by phase, of each pulse corresponding to the MP, the IP and the off-pulse (OFF pulse). See Figure \ref{fig:pulseprofile}. Our analysis subtracts OFF pulse contributions from both MP and IP in order to eliminate the influence from the Crab Nebula.  Details follow. 

The MP window width is 0.2 of the phase, from 0.9 to 1.1 in phase. It has a background window of equal width, from 0.6 in the phase to 0.8 in the phase. The IP window is 0.15 of the phase wide, from 0.3 in the phase to 0.45 in the phase. We use a background window, ``OFF",  of equal width, from 0.6 in the phase to 0.75 in the phase. The ON and OFF windows are separated by  0.1 in phase to avoid accidentally including part of either pulse in the background window. 

$N_{MP}$ and $N_{IP}$ are the number of photons received in either MP or IP ON window respectively.  $N_{MP, OFF}$ and $N_{IP, OFF}$ are the number of photons received in the OFF windows respectively. 
We calculate differences as shown below in order to obtain total photon counts in only the pulsed emission from the pulsar. $M$ and $I$ are the pulsed emission photon counts for the MP and IP respectively.

\begin{equation}
    M  = N_{MP} - N_{MP, OFF}
\end{equation}

\begin{equation}
    I  = N_{IP}- N_{IP, OFF}
\end{equation} 
So, the photon-count data is written as:
\begin{equation}
X_M = \{M_{1}, M_{2}, M_{3}... M_{m}\} 
\end{equation}
for the MP and
\begin{equation}
X_I = \{I_{1}, I_{2}, I_{3}... I_{m}\}
\end{equation}
for the IP
where $m = 2122264$, which is the number of pulses of usable Crab data. 

To create the distribution that results from adding each pair of successive pulses together, we complete the following process, creating a new distribution as follows:

\begin{equation}
    \begin{aligned}
    X' = \{ M_{1}+  M_{2},
     M_{3}+ M_{4} \\+...  M_{n-1}+ M_{{n}}\}
    \end{aligned}
    \label{2puleq}
\end{equation}

\begin{figure}[ht!]
\centering
\includegraphics[scale=0.44,angle=0]{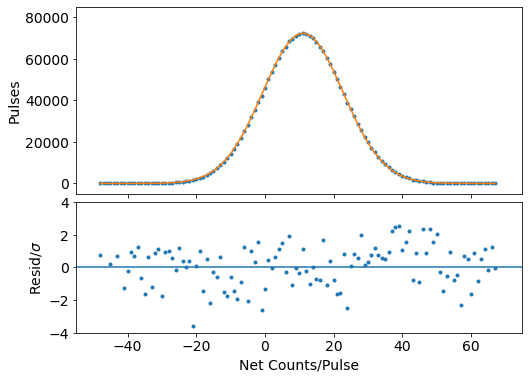} 
\caption{Top panel: the single‑pulse photon‑count distribution for the MP (blue) and the fitted Skellam (orange). Bottom panel:  the difference between the data and model, divided by the uncertainty in the data. Error bars are not shown, as they are smaller than the plotting symbols.}
\label{fig:SkellamFitAndResiduals}
\end{figure}

\begin{figure}[ht!]
\centering
\includegraphics[scale=.44,angle=0]{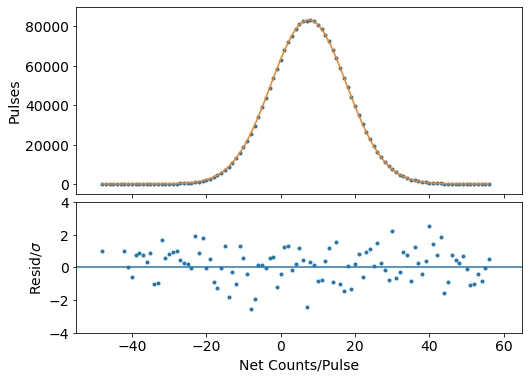} 
\caption{Top panel: the single‑pulse photon‑count distribution for the IP (blue) and the fitted Skellam (orange). Bottom panel: the difference between the data and model, divided by the uncertainty in the data. Error bars are not shown, as they are smaller than the plotting symbols.}
\label{fig:InterPulseOnePulsePerProfileSkellamFit}
\end{figure}

The photon-count distributions shown in Figures \ref{fig:SkellamFitAndResiduals} and \ref{fig:InterPulseOnePulsePerProfileSkellamFit} are histograms of $X_M$ and $X_I$ respectively. 
Throughout the rest of the paper the number of incidences of a certain value $X$ in a particular bin is called $f_i$ where $i$ the bin number. 

\section{Modeling the Photon-Count Distributions}
\label{sec:datanalysis}

\subsection{The Skellam Distribution}

We present a model of the photon-count distribution of the pulses in the Crab pulsar in the X--ray energies observed by the NICER instrument.
We express the measurements in terms of photon counts rather than flux to take full advantage of the underlying Poisson counting statistics. Both the pulsar and the background are individually Poisson‑distributed processes. The difference of two Poisson-distributed photon counts follows a Skellam distribution
\citep{Skellam1946}, which we use to model the difference between the on-pulse and off-pulse counts.

A Poisson distribution is described by the following equation:
\begin{equation}
f(k;\mu)=\frac{\mu^k e^{-\mu}}{k!}
\end{equation}
\noindent where $f$ is the probability that a discrete random variable will be equal to $k$, the number of occurrences. 

The Skellam distribution is a discrete probability distribution defined by the difference of two statistically independent Poisson random variables, ${N_{1}}-{N_{2}}$ with the corresponding means of ${\mu_{1}}$ and ${\mu_{2}}$. It is unimodal and asymmetric, and the values far from the center are generally higher than they are in a normal distribution. 
The Skellam is described by: 
\begin{equation}
{\displaystyle P(k;\mu_1,\mu_2)=e^{-(\mu _{1} \!+\! \mu _{2})}\left({\frac {\mu _{1}}{\mu _{2}}}\right)^{k/2}\!\!I_{k}(2{\sqrt {\mu _{1}\mu _{2}}})}
\label{skellameq}
\end{equation}
where \textit{I}\textsubscript{\textit{k}} is the modified Bessel function.

\subsection{Statistics used to choose models}
We find a best-fit Gaussian and best-fit Skellam using least-squares minimization. For each of the two models, we used three methods for assessing and comparing the quality of the model: $\chi^2_\nu$,  Akaike Information Criterion (AIC), and Bayesian Information Criterion (BIC)\citep{Akaike1981, SchwarzBIC1978}.  We briefly describe each statistic below.  

We compute $\chi^2_{\nu}$ using:

\begin{eqnarray}
    \chi^2_\nu = \frac{1}{N-A}\Sigma \frac{(f_i -m_i)^2}{f_i}
\label{chieqn}
\end{eqnarray}
where $f_i$ is the number of photon-count measurements in bin $i$, and $m_i$ is the expected number of photon-count measurements in that bin according to a model. $f_i$ is in the denominator as the square of the uncertainty on $f_i$ which we take to be Poisson distributed.  $A$ is the number of parameters and $N$ is the number of bins in the distribution of photon counts.

The AIC depends upon the log likelihood which is computed via the likelihood, $L(\theta)$:

\begin{equation}
    L(\theta) = \prod_{i=0}^{N} \frac{1}{\sqrt{2\pi f_i}} \exp\left(-\frac{(f_i - m_i)^2}{2x_i}\right)
\end{equation}

which gives log likelihood $\ell(\theta) as$
\begin{equation}
\label{likelihood_big}
\ell(\theta) = \sum_{i=0}^{N} \left[ -\frac{1}{2}\log(2\pi f_i) - \frac{(f_i - m_i)^2}{2x_i} \right]
\end{equation}

In order to show the relationship between likelihood and $\chi^2$ 
we separate the log likelihood into two sums.
\[
\ell(\theta) = -\frac{1}{2}\left[\sum_{i=0}^{N}  \log(2\pi f_i) + \sum_{i=0}^{N}\frac{(f_i - m_i)^2}{f_i}\right] 
\]
We express this in terms of $C$, a term that does not depend upon the model, and $\chi^2$ which does:
We define
\begin{eqnarray}
    \label{Cdef}
    C \equiv \sum_{i=0}^{N}  \log(2\pi f_i)
\end{eqnarray}
Then we can express the log likelihood as
\begin{eqnarray}
    \label{loglike}
    \ell(\theta) =- \frac{1}{2} (C + \chi^2)
\end{eqnarray}

AIC is given by
    \begin{eqnarray}
        \text{AIC} &=& 2A - 2\ell(\hat{\theta})\\
        &=& 2A + C + \chi^2
    \end{eqnarray}
and the BIC is given by 
\begin{eqnarray}
        \text{BIC} &=& A\log(N) - 2\ell(\hat{\theta})\\
        &=& A\log(N) + C + \chi^2
    \end{eqnarray}

The AIC and BIC are essentially $\chi^2$ plus a penalty for the number of parameters. $\Delta AIC > 2 $ or $\Delta BIC > 2$ indicates preference for a model\citep{Akaike1981,SchwarzBIC1978}.

\section{Results}
\label{sec:Measuring the Counts and Binning the Data}
\subsection{One Pulse per Profile}
\label{sec:One Pulse}
Table \ref{table:stats single-pulse main} shows the $\chi^2_\nu$, AIC, and BIC results for the MP single-pulse photon-count distribution and Table \ref{table:stats single-pulse interpulse} shows the same for the IP. We also show the change in the statistic value in each metric $\Delta$ and $\ell(\theta)$ to allow the reader to replicate calculations. 
All three metrics suggest that the Skellam is the more appropriate model for both the MP and the IP. 

In both the AIC and the BIC a difference of two between two models represents a preference, and a difference of  30--40, as we obtain, represents an overwhelming preference \citep{Burnham2002}.  

We begin our discussion with the IP, because the reduced $\chi^2$ (hereafter $\chi^2_\nu$) of 1.037 for 98 degrees of freedom indicates an excellent fit, so we are confident that using the Skellam to model the data is appropriate. The $\chi_\nu^2$ of the Gaussian model is 1.716 with 97 degrees of freedom, and likelihood ratio of the Skellam over the Gaussian is 2300.

The corresponding reduced $\chi^2_\nu$ for the Skellam model of the MP is surprisingly high at 1.57 with 110 degrees of freedom, corresponding to a $\sim4\sigma$ deviation from the expected value and indicating that additional physics is required to describe the MP variability. The reduced $\chi^2_\nu$ of the Gaussian model is even higher, at 2.37 with 109 degrees of freedom. The ratio of the $\chi^2$ tail probabilities for these two models is $\approx 3 \times 10^{9}$, indicating that while the data are very unlikely to be described by a pure Skellam distribution, they are far less likely to be described by a Gaussian.

Given the validating high-probability of the Skellam model for the IP, we interpret our results for the MP to be a Skellam distribution plus some unmodeled excess variability. 

Figures \ref{fig:SkellamFitAndResiduals} and \ref{fig:InterPulseOnePulsePerProfileSkellamFit} show the data, the Skellam fit,and the residuals per $\sigma$ for the MP and IP respectively. 
The figures give us more insight into the excess variability observed in the MP. They show that the residuals per $\sigma$ are evenly distributed for the IP, but not for the MP. The MP's residuals per $\sigma$ are higher at higher photon counts, suggesting that high-end tail of the distribution is the location of disagreement. We can see that the IP's residuals do not display similar behavior, remaining consistent across all photon counts. 

\begin{table} [h!]
\begin{center}
\begin{tabular}{| c|c|c|c |}
\hline
  &Skellam& Gaussian& $\Delta$ (Gaussian-Skellam)\\
 \hline
 $\chi^2_\nu$&1.57 &  2.37 & 0.81\\
 \hline
 AIC&608.5&653.1& 44.6\\
 \hline
 BIC&614.0& 656.6& 42.6\\
 \hline
 $\ell(\theta)$&-604.5 & -647.1 & -42.6\\\hline
\end{tabular}
\caption{$\chi^2$, AIC, and BIC values for MP single-pulse photon‑count distributions for each model (Gaussian and Skellam) and the difference in the metrics $\Delta$ between the two models. The log‑likelihood function $\ell(\theta)$ is included to allow readers to reproduce the calculations (Equations \ref{likelihood_big} and \ref{loglike}). All three metrics suggest the Skellam model is preferred.}
\label{table:stats single-pulse main}
\end{center}
\end{table}

\begin{table}
\begin{center}
\begin{tabular}{| c|c|c|c |}
\hline
  &Skellam& Gaussian& $\Delta$ (Gaussian-Skellam)\\
 \hline
 $\chi^2_\nu$&1.037&  1.716& 0.679\\
 \hline
 AIC    &514.9  &549.2  & 34.3\\
 \hline
 BIC    &520.2  & 552.6 & 32.4\\
 \hline
$\ell(\theta)$&-510.9& -543.2& -32.3\\\hline
\end{tabular}
\caption{$\chi^2$, AIC, and BIC values for IP single-pulse photon‑count distributions for each model (Gaussian and Skellam) and the difference in the metrics $\Delta$ between the two models. The log‑likelihood function $\ell(\theta)$ is included to allow readers to reproduce the calculations (Equations \ref{likelihood_big} and \ref{loglike}). All three metrics suggest the Skellam model is preferred.}
\label{table:stats single-pulse interpulse}
\end{center}
\end{table}

\subsection{Memory}
\label{sec:Two Pulses}

\begin{figure}[ht]
\centering
\includegraphics[scale=0.44,angle=0]{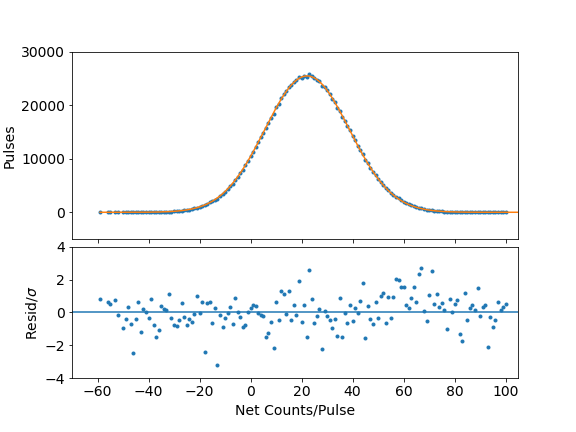} 
\caption{Top panel:  the 2‑pulse photon‑count distribution for the MP (blue) and the fitted Skellam (orange). Bottom panel: residuals/uncertainty. Error bars are not shown, as they are smaller than the plotting symbols..}
\label{fig:TwoPulseAndResiduals}
\end{figure}

\begin{figure}[ht]
\centering
\includegraphics[scale=0.44,angle=0]{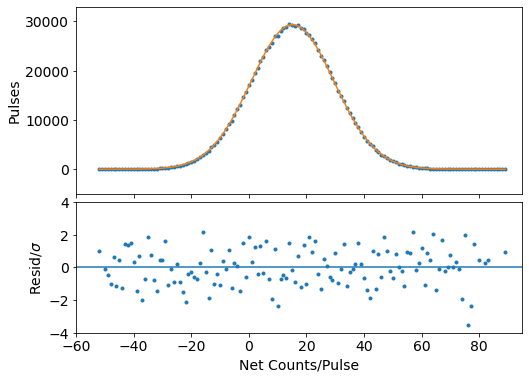} 
\caption{Top panel: shows the 2‑pulse photon‑count distribution for the IP (blue) and the fitted Skellam (orange). Bottom panel: residuals/uncertainty. Error bars are not shown, as they are smaller than the plotting symbols.}
\label{fig:InterpulseTwoPulseAndResiduals}
\end{figure}

If a particular pulse shows excess variability, is the following pulse more likely to show excess variability?  To examine this `memory' effect we conducted two different analyses. First, we repeated the photon-count measurements on data containing two pulses per profile, instead of just one (see Equation \ref{2puleq}). 
If the single-pulse photon-count following a large pulse is more likely to also be large, then this analysis will produce an even higher $\chi_\nu^2$ for the two pulse data than the single pulse data. However, if there are no such memory processes, we would expect to see a lower $\chi^2$ for the two pulse data when compared with the single pulse data. The fits and residuals for the MP can be seen in Figure \ref{fig:TwoPulseAndResiduals} and the IP fits and residuals can be seen in Figure \ref{fig:InterpulseTwoPulseAndResiduals}.  Our results, shown in Table \ref{table: MP and IP one and two successive pulses comparison}, are not suggestive of a memory process, in fact $\chi_\nu^2=1.006$ for the MP, suggesting that once neighboring pulses are added together the excess variability is undetectable.

The autocorrelation of the photon-count data $X$ for the MP up to 30 lags to also tests whether the photon-count on pulse is related to the photon-count of subsequent pulses. For the MP, the lag-1 autocorrelation coefficient is $r_1 = 0.002$, with an expected sampling uncertainty of $\sigma_r \approx N^{-1/2} \simeq 7\times10^{-4}$ (which is very similar to the lag-1 coefficient when we randomly shuffle the pulses). Here $\rho_1 = \mathrm{Corr}(X_n, X_{n+1})$ denotes the intrinsic lag-1 correlation coefficient between successive pulse photon counts, quantifying the degree to which the variability of one pulse is linearly related to that of the immediately preceding pulse. This corresponds to a 95\% confidence upper limit of $\rho_1 < 3.4 \times 10^{-3}$ on any pulse-to-pulse correlation. Although such a correlation is marginally detectable given the very large sample size, its magnitude is extremely small and the autocorrelation falls rapidly at higher lags (e.g., $r_2 \sim 10^{-4}$). Furthermore, the $\chi_\nu^2$ of two-pulse profiles is lower than that of single-pulse profiles, as expected for statistically independent pulses. Together, these results rule out any physically meaningful pulse-to-pulse memory in the MP X--ray emission.

 \begin{deluxetable}{|c| c |c |c| c|}
\tablecolumns{5}
\label{table: MP and IP one and two successive pulses comparison} 
\tablehead{
& \multicolumn{2}{c|}{1-pulse} & \multicolumn{2}{c|}{2-pulse}
\\
Skellam  & MP  & IP  & MP  & IP} 
\startdata
  $\chi^2_\nu$  &  1.570 & 1.037 & 1.006 & 1.070\\
  $\ell(\theta)$ & -604.5 & -510.9 & -727.3 & -640.2\\
  AIC            & 608.5 &  514.9 & 731.3 & 644.2 \\
  BIC            & 614.0 & 520.1 & 737.5 & 650.03 \\
\enddata
 \caption {$\chi^2$, AIC, and BIC values for each photon‑count distribution (one‑pulse and two‑pulse) using the Skellam model. The log‑likelihood function $\ell(\theta)$ is included to allow readers to reproduce the calculations (Equations \ref{likelihood_big} and \ref{loglike}). Both the MP and IP datasets show stronger agreement with the Skellam model for the single‑pulse distributions.}
\end{deluxetable}

\subsection{Alternate Pulse Windows}
\label{sec:Differences in Windows}

As GRPs occur slightly later in rotational phase than the X--ray pulses, any influence associated with GRPs would be expected to preferentially affect the trailing part of the X--ray main pulse. To investigate this possibility, we divided the MP phase window into two equal halves and constructed single--pulse photon--count histograms for each half, which were fit with the Skellam distribution. The resulting $\chi^2_\nu$ values are 0.99 for the leading half and 1.13 for the trailing half, suggesting that the leading portion of the MP is well described by the Skellam model, while the trailing half exhibits modest excess variability. We note, however, that splitting the MP window reduces the mean photon count per pulse in each half from $\sim11$ to $\sim5$, placing the analysis in a low--count regime where Poisson discreteness becomes significant and $\chi^2$--based goodness--of--fit tests are less reliable. In this regime, small deviations in bin occupancy can inflate $\chi^2_\nu$ even when the underlying distribution is correctly specified. Consequently, while the difference between the two halves is suggestive, it should be interpreted with caution.

We also repeated the analysis using phase windows half the width of the original ones, while keeping the window centers fixed. For this test, the MP window spanned 0.1 in phase (0.95--1.05), and the IP window spanned 0.075 in phase (0.3375--0.4125). The resulting $\chi^2_\nu$ values are 1.34 for the MP and 1.13 for the IP, indicating increased apparent scatter relative to the Skellam model in both cases, with a stronger effect observed for the MP. As with the half--pulse analysis, narrowing the phase windows substantially reduces the mean photon counts per pulse, to $\sim8$ for the MP and $\sim5$ for the IP. At these low count levels, the performance of histogram--based $\chi^2$ tests degrades, and modest increases in $\chi^2_\nu$ are expected even for a correct underlying distribution. In the limit of very narrow phase windows, where individual pulses contain only a few photons, goodness--of--fit tests become increasingly insensitive and eventually ill--defined. We therefore interpret the narrow--window results as reflecting reduced statistical power and enhanced sensitivity to small--number fluctuations, rather than as evidence against the Skellam description. This interpretation is consistent with the full--window one--pulse and two--pulse analyses, which show that the interpulse is very well described by a Skellam distribution when sufficient photon statistics are retained.

\section{Discussion}
\label{sec:results}

Our phase--resolved photon--count analysis reveals a clear divergence in statistical behavior between the Crab pulsar’s main pulse (MP) and interpulse (IP). The IP is well described by a Skellam distribution, consistent with a Poissonian signal superposed on Poisson background, whereas the MP exhibits a significant excess of variability on a pulse--by--pulse basis, driven primarily by high--count events. When photon counts are summed over two successive pulses, however, the MP distribution becomes consistent with Skellam expectations, with $ \chi^2_\nu \approx 1.006$. This indicates that the excess variability does not persist across rotations and instead averages out rapidly when successive pulses are combined. Consistent with this interpretation, our search for short--lag (``memory'') correlations finds no significant lag--1 autocorrelation, arguing against rotation--to--rotation persistence as the source of the MP excess. Collectively, these results imply that the enhanced variability in the MP arises from short--lived, pulse--localized processes, while the IP remains well described by a purely Poissonian model.

We briefly explored potential connections between the MP excess variability and giant radio pulses (GRPs). Previous work has shown that X--ray emission coincident with GRPs is enhanced by $3.8 \pm 0.7\%$ in the MP phase window \citep{Enoto2021}, but GRPs occur in only a small fraction of pulsar rotations. When averaged over all pulses, the implied increase in the mean MP count rate is therefore at the $\sim 10^{-4}$ level. Simple simulations in which excess counts were added to reproduce this level of enhancement did not yield a detectable increase in the photon--count variance as measured by $\chi^2$. While these tests are necessarily approximate, they suggest that the observed MP excess variability cannot be explained solely by GRP--associated X--ray enhancements. A more detailed treatment of this connection is left for future work. 

While we do not attempt to provide a comprehensive or prescriptive treatment of Skellam statistics here, our analysis does suggest a simple and practical rule of thumb. When the mean values of two Poisson variates, $\langle x\rangle$ and $\langle y\rangle$, are comparable—specifically when $\langle x\rangle$ exceeds $\langle y\rangle$ but not by a large factor—and when the sample size is sufficiently large that the difference $z = x -y$ can take on negative values, the Skellam distribution is the appropriate statistical description. The NICER IP results demonstrate this clearly: Gaussian approximations perform poorly, Skellam provides an excellent description, and modified Poisson approaches fail outright because they cannot accommodate negative differences. Further exploration of Skellam--based methods is beyond the scope of this paper, but our results suggest that they may be broadly useful in astrophysical contexts involving differenced Poisson data, particularly where large photon statistics enable sensitive tests of distributional assumptions.

\begin{acknowledgments}
Portions of this research performed at NRL and at Haverford College were supported by NASA. This project has made use of data products and software provided by the High Energy Astrophysics Science Archive Research Center (HEASARC), which is a service of the Astrophysics Science Division at NASA/GSFC and the High Energy Astrophysics Division of the Smithsonian Astrophysical Observatory. The research presented has relied on NASA's Astrophysics Data System (ADS) bibliographic services and the ArXiv.
\end{acknowledgments}

\facilities{NICER \citep{Gendreau2017}}
\software{HEASoft \citep{HEASarc}, PINT \citep{Luo_2021}}

\bibliography{myreferences}

@software{HEASarc,
       author = {{Nasa High Energy Astrophysics Science Archive Research Center (Heasarc)}},
        title = "{HEAsoft: Unified Release of FTOOLS and XANADU}",
 howpublished = {Astrophysics Source Code Library, record ascl:1408.004},
         year = 2014,
        month = aug,
          eid = {ascl:1408.004},
archivePrefix = {ascl},
       eprint = {1408.004},
       adsurl = {https://ui.adsabs.harvard.edu/abs/2014ascl.soft08004N}
}

@article{Luo_2021,
	author = {Luo, Jing and Ransom, Scott and Demorest, Paul and Ray, Paul S. and Archibald, Anne and Kerr, Matthew and Jennings, Ross J. and Bachetti, Matteo and van Haasteren, Rutger and Champagne, Chloe A. and Colen, Jonathan and Phillips, Camryn and Zimmerman, Josef and Stovall, Kevin and Lam, Michael T. and Jenet, Fredrick A.},
	doi = {10.3847/1538-4357/abe62f},
	journal = {The Astrophysical Journal},
	month = {apr},
	number = {1},
	pages = {45},
	publisher = {The American Astronomical Society},
	title = {PINT: A Modern Software Package for Pulsar Timing},
	url = {https://doi.org/10.3847/1538-4357/abe62f},
	volume = {911},
	year = {2021}}

@ARTICLE{SchwarzBIC1978,
       author = {{Schwarz}, Gideon},
        title = "{Estimating the Dimension of a Model}",
      journal = {Annals of Statistics},
         year = 1978,
        month = jul,
       volume = {6},
       number = {2},
        pages = {461-464},
       adsurl = {https://ui.adsabs.harvard.edu/abs/1978AnSta...6..461S}
}

@article{Akaike1981,
	author = {Hirotugu Akaike},
	doi = {https://doi.org/10.1016/0304-4076(81)90071-3},
	issn = {0304-4076},
	journal = {Journal of Econometrics},
	number = {1},
	pages = {3-14},
	title = {Likelihood of a model and information criteria},
	url = {https://www.sciencedirect.com/science/article/pii/0304407681900713},
	volume = {16},
	year = {1981}}

@article{Knight2006,
  author  = {Knight, H. S.},
  title   = {Giant Radio Pulses from the Crab Pulsar},
  journal = {The Astrophysical Journal},
  volume  = {640},
  pages   = {941--950},
  year    = {2006},
  doi     = {10.1086/500123}
}

@article{Cash1979,
  author  = {Cash, W.},
  title   = {Parameter Estimation in Astronomy through Application of the Likelihood Ratio},
  journal = {The Astrophysical Journal},
  volume  = {228},
  pages   = {939--947},
  year    = {1979},
  doi     = {10.1086/156922}
}

@article{BakerCousins1984,
  author  = {Baker, S. and Cousins, R. D.},
  title   = {Clarification of the Use of {\\ensuremath{\\chi^2}} and Likelihood Functions in Fits to Histograms},
  journal = {Nuclear Instruments and Methods in Physics Research},
  volume  = {221},
  pages   = {437--442},
  year    = {1984},
  doi     = {10.1016/0167-5087(84)90016-4}
}

@article{Gehrels1986,
  author  = {Gehrels, N.},
  title   = {Confidence Limits for Small Numbers of Events in Astrophysical Data},
  journal = {The Astrophysical Journal},
  volume  = {303},
  pages   = {336--346},
  year    = {1986},
  doi     = {10.1086/164079}
}

@article{Mighell1999,
  author  = {Mighell, K. J.},
  title   = {Parameter Estimation in Astronomy with Poisson-Distributed Data},
  journal = {The Astrophysical Journal},
  volume  = {518},
  pages   = {380--393},
  year    = {1999},
  doi     = {10.1086/307255}
}

@book{Burnham2002,
	author = {Burnham, K.P. and Anderson, D.R.},
	publisher = {New York: Springer-Verlag},
	title = {Model Selection and Multimodel Inference: A Practical Information-Theoretic Approach},
	year = {2002}}

@book{abramowitz1964handbook,
  title     = {Handbook of Mathematical Functions with Formulas, Graphs, and Mathematical Tables},
  author    = {Abramowitz, Milton and Stegun, Irene A.},
  year      = {1964},
  publisher = {National Bureau of Standards},
  address   = {Washington, D.C.},
  series    = {Applied Mathematics Series},
  number    = {55}
}

@ARTICLE{Lundgren95,
       author = {{Lundgren}, S.~C. and {Cordes}, J.~M. and {Ulmer}, M. and {Matz}, S.~M. and
         {Lomatch}, S. and {Foster}, R.~S. and {Hankins}, T.},
        title = "{Giant Pulses from the Crab Pulsar: A Joint Radio and Gamma-Ray Study}",
      journal = {\apj},
         year = 1995,
        month = nov,
       volume = {453},
        pages = {433},
          doi = {10.1086/176404},
       adsurl = {https://ui.adsabs.harvard.edu/abs/1995ApJ...453..433L}
}

@ARTICLE{Deneva19,
       author = {{Deneva}, J.~S. and {Ray}, P.~S. and {Lommen}, A. and {Ransom}, S.~M. and
         {Bogdanov}, S. and {Kerr}, M. and {Wood}, K.~S. and {Arzoumanian}, Z. and
         {Black}, K. and {Doty}, J. and {Gendreau}, K.~C. and {Guillot}, S. and
         {Harding}, A. and {Lewandowska}, N. and {Malacaria}, C. and
         {Markwardt}, C.~B. and {Price}, S. and {Winternitz}, L. and
         {Wolff}, M.~T. and {Guillemot}, L. and {Cognard}, I. and
         {Baker}, P.~T. and {Blumer}, H. and {Brook}, P.~R. and
         {Cromartie}, H.~T. and {Demorest}, P.~B. and {DeCesar}, M.~E. and
         {Dolch}, T. and {Ellis}, J.~A. and {Ferdman}, R.~D. and
         {Ferrara}, E.~C. and {Fonseca}, E. and {Garver-Daniels}, N. and
         {Gentile}, P.~A. and {Jones}, M.~L. and {Lam}, M.~T. and
         {Lorimer}, D.~R. and {Lynch}, R.~S. and {McLaughlin}, M.~A. and
         {Ng}, C. and {Nice}, D.~J. and {Pennucci}, T.~T. and {Spiewak}, R. and
         {Stairs}, I.~H. and {Stovall}, K. and {Swiggum}, J.~K. and {Vigeland
        }, S.~J. and {Zhu}, W.~W.},
        title = "{High-precision X-Ray Timing of Three Millisecond Pulsars with NICER: Stability Estimates and Comparison with Radio}",
      journal = {\apj},
         year = 2019,
        month = apr,
       volume = {874},
       number = {2},
          eid = {160},
        pages = {160},
          doi = {10.3847/1538-4357/ab0966},
archivePrefix = {arXiv},
       eprint = {1902.07130},
 primaryClass = {astro-ph.HE},
       adsurl = {https://ui.adsabs.harvard.edu/abs/2019ApJ...874..160D}
}

@ARTICLE{Hankins2016,
       author = {{Hankins}, T.~H. and {Eilek}, J.~A. and {Jones}, G.},
        title = "{The Crab Pulsar at Centimeter Wavelengths. II. Single Pulses}",
      journal = {\apj},
         year = 2016,
        month = dec,
       volume = {833},
       number = {1},
          eid = {47},
        pages = {47},
          doi = {10.3847/1538-4357/833/1/47},
archivePrefix = {arXiv},
       eprint = {1608.08881},
 primaryClass = {astro-ph.HE},
       adsurl = {https://ui.adsabs.harvard.edu/abs/2016ApJ...833...47H}
}

@ARTICLE{Hankins2007,
       author = {{Hankins}, T.~H. and {Eilek}, J.~A.},
        title = "{Radio Emission Signatures in the Crab Pulsar}",
      journal = {\apj},
         year = 2007,
        month = nov,
       volume = {670},
       number = {1},
        pages = {693-701},
          doi = {10.1086/522362},
archivePrefix = {arXiv},
       eprint = {0708.2505},
 primaryClass = {astro-ph},
       adsurl = {https://ui.adsabs.harvard.edu/abs/2007ApJ...670..693H}
}

@ARTICLE{Abdo2010,
       author = {{Abdo}, A.~A. and {Ackermann}, M. and {Ajello}, M. and {Atwood}, W.~B. and {Axelsson}, M. and {Baldini}, L. and {Ballet}, J. and {Barbiellini}, G. and {Baring}, M.~G. and {Bastieri}, D. and {Bechtol}, K. and {Bellazzini}, R. and {Berenji}, B. and {Blandford}, R.~D. and {Bloom}, E.~D. and {Bonamente}, E. and {Borgland}, A.~W. and {Bregeon}, J. and {Brez}, A. and {Brigida}, M. and {Bruel}, P. and {Burnett}, T.~H. and {Caliandro}, G.~A. and {Cameron}, R.~A. and {Camilo}, F. and {Caraveo}, P.~A. and {Casandjian}, J.~M. and {Cecchi}, C. and {{\c{C}}elik}, {\"O}. and {Chekhtman}, A. and {Cheung}, C.~C. and {Chiang}, J. and {Ciprini}, S. and {Claus}, R. and {Cognard}, I. and {Cohen-Tanugi}, J. and {Cominsky}, L.~R. and {Conrad}, J. and {Dermer}, C.~D. and {de Angelis}, A. and {de Luca}, A. and {de Palma}, F. and {Digel}, S.~W. and {Silva}, E. do Couto e. and {Drell}, P.~S. and {Dubois}, R. and {Dumora}, D. and {Espinoza}, C. and {Farnier}, C. and {Favuzzi}, C. and {Fegan}, S.~J. and {Ferrara}, E.~C. and {Focke}, W.~B. and {Frailis}, M. and {Freire}, P.~C.~C. and {Fukazawa}, Y. and {Funk}, S. and {Fusco}, P. and {Gargano}, F. and {Gasparrini}, D. and {Gehrels}, N. and {Germani}, S. and {Giavitto}, G. and {Giebels}, B. and {Giglietto}, N. and {Giordano}, F. and {Glanzman}, T. and {Godfrey}, G. and {Grenier}, I.~A. and {Grondin}, M. -H. and {Grove}, J.~E. and {Guillemot}, L. and {Guiriec}, S. and {Hanabata}, Y. and {Harding}, A.~K. and {Hayashida}, M. and {Hays}, E. and {Hughes}, R.~E. and {J{\'o}hannesson}, G. and {Johnson}, A.~S. and {Johnson}, R.~P. and {Johnson}, T.~J. and {Johnson}, W.~N. and {Johnston}, S. and {Kamae}, T. and {Katagiri}, H. and {Kataoka}, J. and {Kawai}, N. and {Kerr}, M. and {Kn{\"o}dlseder}, J. and {Kocian}, M.~L. and {Kramer}, M. and {Kuehn}, F. and {Kuss}, M. and {Lande}, J. and {Latronico}, L. and {Lee}, S. -H. and {Lemoine-Goumard}, M. and {Longo}, F. and {Loparco}, F. and {Lott}, B. and {Lovellette}, M.~N. and {Lubrano}, P. and {Lyne}, A.~G. and {Makeev}, A. and {Marelli}, M. and {Mazziotta}, M.~N. and {McEnery}, J.~E. and {Meurer}, C. and {Michelson}, P.~F. and {Mitthumsiri}, W. and {Mizuno}, T. and {Moiseev}, A.~A. and {Monte}, C. and {Monzani}, M.~E. and {Moretti}, E. and {Morselli}, A. and {Moskalenko}, I.~V. and {Murgia}, S. and {Nakamori}, T. and {Nolan}, P.~L. and {Norris}, J.~P. and {Noutsos}, A. and {Nuss}, E. and {Ohsugi}, T. and {Omodei}, N. and {Orlando}, E. and {Ormes}, J.~F. and {Ozaki}, M. and {Paneque}, D. and {Panetta}, J.~H. and {Parent}, D. and {Pelassa}, V. and {Pepe}, M. and {Pesce-Rollins}, M. and {Pierbattista}, M. and {Piron}, F. and {Porter}, T.~A. and {Rain{\`o}}, S. and {Rando}, R. and {Ray}, P.~S. and {Razzano}, M. and {Reimer}, A. and {Reimer}, O. and {Reposeur}, T. and {Ritz}, S. and {Rochester}, L.~S. and {Rodriguez}, A.~Y. and {Romani}, R.~W. and {Roth}, M. and {Ryde}, F. and {Sadrozinski}, H.~F. -W. and {Sanchez}, D. and {Sander}, A. and {Saz Parkinson}, P.~M. and {Scargle}, J.~D. and {Sgr{\`o}}, C. and {Siskind}, E.~J. and {Smith}, D.~A. and {Smith}, P.~D. and {Spandre}, G. and {Spinelli}, P. and {Stappers}, B.~W. and {Strickman}, M.~S. and {Suson}, D.~J. and {Tajima}, H. and {Takahashi}, H. and {Tanaka}, T. and {Thayer}, J.~B. and {Thayer}, J.~G. and {Theureau}, G. and {Thompson}, D.~J. and {Thorsett}, S.~E. and {Tibaldo}, L. and {Torres}, D.~F. and {Tosti}, G. and {Tramacere}, A. and {Uchiyama}, Y. and {Usher}, T.~L. and {Van Etten}, A. and {Vasileiou}, V. and {Vilchez}, N. and {Vitale}, V. and {Waite}, A.~P. and {Wallace}, E. and {Wang}, P. and {Watters}, K. and {Weltevrede}, P. and {Winer}, B.~L. and {Wood}, K.~S. and {Ylinen}, T. and {Ziegler}, M.},
        title = "{Fermi Large Area Telescope Observations of the Crab Pulsar And Nebula}",
      journal = {\apj},
         year = 2010,
        month = jan,
       volume = {708},
       number = {2},
        pages = {1254-1267},
          doi = {10.1088/0004-637X/708/2/1254},
archivePrefix = {arXiv},
       eprint = {0911.2412},
 primaryClass = {astro-ph.HE},
       adsurl = {https://ui.adsabs.harvard.edu/abs/2010ApJ...708.1254A}
}

@ARTICLE{Rowan2020,
       author = {{Rowan}, Dominick M. and {Ghazi}, Zaynab and {Lugo}, Lauren and {Spano}, Elizabeth and {Lommen}, Andrea and {Harding}, Alice and {Venter}, Christo and {Ludlam}, Renee and {Ray}, Paul S. and {Kerr}, Matthew and {Arzoumanian}, Zaven and {Bogdanov}, Slavko and {Deneva}, Julia and {Guillot}, Sebastien and {Lewandowska}, Natalia and {Markwardt}, Craig B. and {Ransom}, Scott and {Enoto}, Teruaki and {Wood}, Kent S. and {Gendreau}, Keith C.},
        title = "{A NICER View of Spectral and Profile Evolution for Three X-Ray-emitting Millisecond Pulsars}",
      journal = {\apj},
         year = 2020,
        month = apr,
       volume = {892},
       number = {2},
          eid = {150},
        pages = {150},
          doi = {10.3847/1538-4357/ab718f},
archivePrefix = {arXiv},
       eprint = {2001.11513},
 primaryClass = {astro-ph.HE},
       adsurl = {https://ui.adsabs.harvard.edu/abs/2020ApJ...892..150R}
}

@ARTICLE{Gendreau2017,
       author = {{Gendreau}, Keith and {Arzoumanian}, Zaven},
        title = "{Searching for a pulse}",
      journal = {Nature Astronomy},
         year = 2017,
        month = dec,
       volume = {1},
        pages = {895-895},
          doi = {10.1038/s41550-017-0301-3},
       adsurl = {https://ui.adsabs.harvard.edu/abs/2017NatAs...1..895G}
}

@ARTICLE{Enoto2021,
       author = {{Enoto}, Teruaki and {Terasawa}, Toshio and {Kisaka}, Shota and {Hu}, Chin-Ping and {Guillot}, Sebastien and {Lewandowska}, Natalia and {Malacaria}, Christian and {Ray}, Paul S. and {Ho}, Wynn C.~G. and {Harding}, Alice K. and {Okajima}, Takashi and {Arzoumanian}, Zaven and {Gendreau}, Keith C. and {Wadiasingh}, Zorawar and {Markwardt}, Craig B. and {Soong}, Yang and {Kenyon}, Steve and {Bogdanov}, Slavko and {Majid}, Walid A. and {G{\"u}ver}, Tolga and {Jaisawal}, Gaurava K. and {Foster}, Rick and {Murata}, Yasuhiro and {Takeuchi}, Hiroshi and {Takefuji}, Kazuhiro and {Sekido}, Mamoru and {Yonekura}, Yoshinori and {Misawa}, Hiroaki and {Tsuchiya}, Fuminori and {Aoki}, Takahiko and {Tokumaru}, Munetoshi and {Honma}, Mareki and {Kameya}, Osamu and {Oyama}, Tomoaki and {Asano}, Katsuaki and {Shibata}, Shinpei and {Tanaka}, Shuta J.},
        title = "{Enhanced x-ray emission coinciding with giant radio pulses from the Crab Pulsar}",
      journal = {Science},
         year = 2021,
        month = apr,
       volume = {372},
       number = {6538},
        pages = {187-190},
          doi = {10.1126/science.abd4659},
archivePrefix = {arXiv},
       eprint = {2104.03492},
 primaryClass = {astro-ph.HE},
       adsurl = {https://ui.adsabs.harvard.edu/abs/2021Sci...372..187E}
}

@ARTICLE{Majid2011,
       author = {{Majid}, Walid A. and {Naudet}, Charles J. and {Lowe}, Stephen T. and {Kuiper}, Thomas B.~H.},
        title = "{Statistical Studies of Giant Pulse Emission from the Crab Pulsar}",
      journal = {\apj},
         year = 2011,
        month = nov,
       volume = {741},
       number = {1},
          eid = {53},
        pages = {53},
          doi = {10.1088/0004-637X/741/1/53},
archivePrefix = {arXiv},
       eprint = {1108.2307},
 primaryClass = {astro-ph.GA},
       adsurl = {https://ui.adsabs.harvard.edu/abs/2011ApJ...741...53M}
}

@ARTICLE{Shearer2003,
       author = {{Shearer}, A. and {Stappers}, B. and {O'Connor}, P. and {Golden}, A. and {Strom}, R. and {Redfern}, M. and {Ryan}, O.},
        title = "{Enhanced Optical Emission During Crab Giant Radio Pulses}",
      journal = {Science},
         year = 2003,
        month = jul,
       volume = {301},
       number = {5632},
        pages = {493-495},
          doi = {10.1126/science.1084919},
archivePrefix = {arXiv},
       eprint = {astro-ph/0308271},
 primaryClass = {astro-ph},
       adsurl = {https://ui.adsabs.harvard.edu/abs/2003Sci...301..493S}
}

@ARTICLE{Strader2013,
       author = {{Strader}, M.~J. and {Johnson}, M.~D. and {Mazin}, B.~A. and {Spiro Jaeger}, G.~V. and {Gwinn}, C.~R. and {Meeker}, S.~R. and {Szypryt}, P. and {van Eyken}, J.~C. and {Marsden}, D. and {O'Brien}, K. and {Walter}, A.~B. and {Ulbricht}, G. and {Stoughton}, C. and {Bumble}, B.},
        title = "{Excess Optical Enhancement Observed with ARCONS for Early Crab Giant Pulses}",
      journal = {\apjl},
         year = 2013,
        month = dec,
       volume = {779},
       number = {1},
          eid = {L12},
        pages = {L12},
          doi = {10.1088/2041-8205/779/1/L12},
archivePrefix = {arXiv},
       eprint = {1309.3270},
 primaryClass = {astro-ph.HE},
       adsurl = {https://ui.adsabs.harvard.edu/abs/2013ApJ...779L..12S}
}

@ARTICLE{Popov2007,
       author = {{Popov}, M.~V. and {Stappers}, B.},
        title = "{Statistical properties of giant pulses from the Crab pulsar}",
      journal = {\aap},
         year = 2007,
        month = aug,
       volume = {470},
       number = {3},
        pages = {1003-1007},
          doi = {10.1051/0004-6361:20066589},
archivePrefix = {arXiv},
       eprint = {0704.1197},
 primaryClass = {astro-ph},
       adsurl = {https://ui.adsabs.harvard.edu/abs/2007A&A...470.1003P}
}

@ARTICLE{Hesse1974,
       author = {{Hesse}, K.~H. and {Wielebinski}, R.},
        title = "{Pulse Intensity Histograms of Pulsars}",
      journal = {\aap},
         year = 1974,
        month = apr,
       volume = {31},
        pages = {409},
       adsurl = {https://ui.adsabs.harvard.edu/abs/1974A&A....31..409H}
}

@ARTICLE{Bhat2008,
       author = {{Bhat}, N.~D. Ramesh and {Tingay}, Steven J. and {Knight}, Haydon S.},
        title = "{Bright Giant Pulses from the Crab Nebula Pulsar: Statistical Properties, Pulse Broadening, and Scattering Due to the Nebula}",
      journal = {\apj},
         year = 2008,
        month = apr,
       volume = {676},
       number = {2},
        pages = {1200-1209},
          doi = {10.1086/528735},
archivePrefix = {arXiv},
       eprint = {0801.0334},
 primaryClass = {astro-ph},
       adsurl = {https://ui.adsabs.harvard.edu/abs/2008ApJ...676.1200B}
}

@ARTICLE{Yan2022,
       author = {{Yan}, Lin-Li and {Tuo}, You-Li and {Ge}, Ming-Yu and {Lu}, Fang-Jun and {Zheng}, Shi-Jie and {Wang}, Ling-Jun},
        title = "{A Study on the X-Ray Pulse Profile and Spectrum of the Crab Pulsar Using NICER and Insight-HXMT's Observations}",
      journal = {\apj},
         year = 2022,
        month = apr,
       volume = {928},
       number = {2},
          eid = {183},
        pages = {183},
          doi = {10.3847/1538-4357/ac581c},
archivePrefix = {arXiv},
       eprint = {2204.05828},
 primaryClass = {astro-ph.HE},
       adsurl = {https://ui.adsabs.harvard.edu/abs/2022ApJ...928..183Y}
}

@INPROCEEDINGS{Prigozhin2016,
       author = {{Prigozhin}, Gregory and {Gendreau}, Keith and {Doty}, John P. and {Foster}, Richard and {Remillard}, Ronald and {Malonis}, Andrew and {LaMarr}, Beverly and {Vezie}, Michael and {Egan}, Mark and {Villasenor}, Jesus and {Arzoumanian}, Zaven and {Baumgartner}, Wayne and {Scholze}, Frank and {Laubis}, Christian and {Krumrey}, Michael and {Huber}, Alan},
        title = "{NICER instrument detector subsystem: description and performance}",
    booktitle = {Space Telescopes and Instrumentation 2016: Ultraviolet to Gamma Ray},
         year = 2016,
       editor = {{den Herder}, Jan-Willem A. and {Takahashi}, Tadayuki and {Bautz}, Marshall},
       series = {Society of Photo-Optical Instrumentation Engineers (SPIE) Conference Series},
       volume = {9905},
        month = jul,
          eid = {99051I},
        pages = {99051I},
          doi = {10.1117/12.2231718},
       adsurl = {https://ui.adsabs.harvard.edu/abs/2016SPIE.9905E..1IP}
}

@ARTICLE{PINT,
       author = {{Luo}, Jing and {Ransom}, Scott and {Demorest}, Paul and {Ray}, Paul S. and {Archibald}, Anne and {Kerr}, Matthew and {Jennings}, Ross J. and {Bachetti}, Matteo and {van Haasteren}, Rutger and {Champagne}, Chloe A. and {Colen}, Jonathan and {Phillips}, Camryn and {Zimmerman}, Josef and {Stovall}, Kevin and {Lam}, Michael T. and {Jenet}, Fredrick A.},
        title = "{PINT: A Modern Software Package for Pulsar Timing}",
      journal = {\apj},
         year = 2021,
        month = apr,
       volume = {911},
       number = {1},
          eid = {45},
        pages = {45},
          doi = {10.3847/1538-4357/abe62f},
archivePrefix = {arXiv},
       eprint = {2012.00074},
 primaryClass = {astro-ph.IM},
       adsurl = {https://ui.adsabs.harvard.edu/abs/2021ApJ...911...45L}
}

@ARTICLE{timingandglitches,
       author = {{Lyne}, A.~G. and {Jordan}, C.~A. and {Graham-Smith}, F. and {Espinoza}, C.~M. and {Stappers}, B.~W. and {Weltevrede}, P.},
        title = "{45 years of rotation of the Crab pulsar}",
      journal = {\mnras},
         year = 2015,
        month = jan,
       volume = {446},
       number = {1},
        pages = {857-864},
          doi = {10.1093/mnras/stu2118},
archivePrefix = {arXiv},
       eprint = {1410.0886},
 primaryClass = {astro-ph.HE},
       adsurl = {https://ui.adsabs.harvard.edu/abs/2015MNRAS.446..857L}
}

@article{parfiles,
    author = {Lyne, A. G. and Pritchard, R. S. and Graham Smith, F.},
    title = {23 years of Crab pulsar rotational history},
    journal = {Monthly Notices of the Royal Astronomical Society},
    volume = {265},
    number = {4},
    pages = {1003-1012},
    year = {1993},
    month = {12},
    issn = {0035-8711},
    doi = {10.1093/mnras/265.4.1003},
    url = {https://doi.org/10.1093/mnras/265.4.1003},
    eprint = {https://academic.oup.com/mnras/article-pdf/265/4/1003/3173877/mnras265-1003.pdf},
}

@ARTICLE{pint1,
       author = {{Luo}, Jing and {Ransom}, Scott and {Demorest}, Paul and {Ray}, Paul S. and {Archibald}, Anne and {Kerr}, Matthew and {Jennings}, Ross J. and {Bachetti}, Matteo and {van Haasteren}, Rutger and {Champagne}, Chloe A. and {Colen}, Jonathan and {Phillips}, Camryn and {Zimmerman}, Josef and {Stovall}, Kevin and {Lam}, Michael T. and {Jenet}, Fredrick A.},
        title = "{PINT: A Modern Software Package for Pulsar Timing}",
      journal = {\apj},
         year = 2021,
        month = apr,
       volume = {911},
       number = {1},
          eid = {45},
        pages = {45},
          doi = {10.3847/1538-4357/abe62f},
archivePrefix = {arXiv},
       eprint = {2012.00074},
 primaryClass = {astro-ph.IM},
       adsurl = {https://ui.adsabs.harvard.edu/abs/2021ApJ...911...45L}
}

@ARTICLE{pint2,
       author = {{Susobhanan}, Abhimanyu and {Kaplan}, David L. and {Archibald}, Anne M. and {Luo}, Jing and {Ray}, Paul S. and {Pennucci}, Timothy T. and {Ransom}, Scott M. and {Agazie}, Gabriella and {Fiore}, William and {Larsen}, Bjorn and {O'Neill}, Patrick and {van Haasteren}, Rutger and {Anumarlapudi}, Akash and {Bachetti}, Matteo and {Bhakta}, Deven and {Champagne}, Chloe A. and {Cromartie}, H. Thankful and {Demorest}, Paul B. and {Jennings}, Ross J. and {Kerr}, Matthew and {Levina}, Sasha and {McEwen}, Alexander and {Shapiro-Albert}, Brent J. and {Swiggum}, Joseph K.},
        title = "{PINT: Maximum-likelihood Estimation of Pulsar Timing Noise Parameters}",
      journal = {\apj},
         year = 2024,
        month = aug,
       volume = {971},
       number = {2},
          eid = {150},
        pages = {150},
          doi = {10.3847/1538-4357/ad59f7},
archivePrefix = {arXiv},
       eprint = {2405.01977},
 primaryClass = {astro-ph.IM},
       adsurl = {https://ui.adsabs.harvard.edu/abs/2024ApJ...971..150S}
}

@ARTICLE{EOSRef,
       author = {{Bogdanov}, Slavko and {Guillot}, Sebastien and {Ray}, Paul S. and {Wolff}, Michael T. and {Chakrabarty}, Deepto and {Ho}, Wynn C.~G. and {Kerr}, Matthew and {Lamb}, Frederick K. and {Lommen}, Andrea and {Ludlam}, Renee M. and {Milburn}, Reilly and {Montano}, Sergio and {Miller}, M. Coleman and {Baub{\"o}ck}, Michi and {{\"O}zel}, Feryal and {Psaltis}, Dimitrios and {Remillard}, Ronald A. and {Riley}, Thomas E. and {Steiner}, James F. and {Strohmayer}, Tod E. and {Watts}, Anna L. and {Wood}, Kent S. and {Zeldes}, Jesse and {Enoto}, Teruaki and {Okajima}, Takashi and {Kellogg}, James W. and {Baker}, Charles and {Markwardt}, Craig B. and {Arzoumanian}, Zaven and {Gendreau}, Keith C.},
        title = "{Constraining the Neutron Star Mass-Radius Relation and Dense Matter Equation of State with NICER. I. The Millisecond Pulsar X-Ray Data Set}",
      journal = {\apjl},
         year = 2019,
        month = dec,
       volume = {887},
       number = {1},
          eid = {L25},
        pages = {L25},
          doi = {10.3847/2041-8213/ab53eb},
archivePrefix = {arXiv},
       eprint = {1912.05706},
 primaryClass = {astro-ph.HE},
       adsurl = {https://ui.adsabs.harvard.edu/abs/2019ApJ...887L..25B}
}

@article{Skellam1946,
  author  = {Skellam, J. G.},
  title   = {The Frequency Distribution of the Difference Between Two Poisson Variates Belonging to Different Populations},
  journal = {Journal of the Royal Statistical Society, Series A (General)},
  volume  = {109},
  number  = {3},
  pages   = {296},
  year    = {1946}
}

\end{document}